\newcommand{\xib}{\mbox{\boldmath$\xi$}}

\newcommand{\dd}{\text{d}}
\newcommand{\rs}{r_{s}}
\newcommand{\rl}{r_{\Lambda}}

\newcommand{\Bi}{B_{i}}
\newcommand{\Bj}{B_{j}}
\newcommand{\Bk}{B_{k}}

\newcommand{\pai}{\partial_{i}}
\newcommand{\paj}{\partial_{j}}
\newcommand{\pak}{\partial_{k}}

\newcommand{\vr}{\textbf{r}}

\newcommand{\pa}{\partial}

\newcommand{\magne}{\textbf{B}}

\newcommand{\lp}{\left(}
\newcommand{\rp}{\right)}

\newcommand{\dtr}{\dd^{3}r} 

\newcommand{\om}{\mathbf{\Omega}}

\newcommand{\maw}{\mathcal{W}}
\newcommand{\mak}{\mathcal{K}}
\newcommand{\mau}{\mathcal{U}}
\newcommand{\mab}{\mathcal{B}}

\newcommand{\mai}{\mathcal{I}}

\newcommand{\mar}{\mathcal{R}}

\newcommand{\maf}{\mathcal{F}}
\newcommand{\mapp}{\mathcal{P}}
\newcommand{\maa}{\mathcal{A}}

\documentclass[10pt,fleqn]{article}
\usepackage[breaklinks]{hyperref}
\usepackage[german,british]{babel}
\usepackage{fancyheadings}
\usepackage{amssymb,amsmath,amsthm}
\usepackage{latexsym,picinpar,amscd,graphpap,pstcol,pst-all}
\usepackage{graphicx}
\DeclareFixedFont{\tr}{OT1}{pnc}{}{}{17pt}
\DeclareFixedFont{\trdos}{OT1}{pnc}{}{}{10pt} 
\begin{document}
\title{\Large {\tr Equilibrium of large astrophysical structures\\ in the Newton-Hooke spacetime}}
\author{{\Large A. Balaguera-Antol\'{\i}nez}\footnote{E-mail: a-balagu@uniandes.edu.co},
{\Large M. Nowakowski}\footnote{E-mail: mnowakos@uniandes.edu.co}\\
{\small \textit{Departamento de F\'{\i}sica, Universidad de los Andes,}}\\ 
{\small \textit{A.A. 4976, Bogot\'a, D.C., Colombia.}}}
\maketitle

\begin{abstract}
Using the scalar and tensor virial equations, the Lane-Emden equation 
expressing the hydrostatic equilibrium and small oscillations around the equilibrium, we show
how the cosmological constant $\Lambda$ affects various astrophysical quantities important
for large matter conglomeration in the universe.
Among others we examine the effect of $\Lambda$ on the polytropic equation of state
for spherically symmetric objects and find non-negligible results in certain realistic cases.
We calculate the angular velocity for non-spherical oblate 
configurations which demonstrates a clear effect of $\Lambda$ on high eccentricity objects.
We show that for oblate as well as prolate ellipsoids the cosmological constant
influences the critical mass and the temperature of the astrophysical object.
These and other results show that the effect of $\Lambda$ is large for
flat astrophysical bodies.
\end{abstract}
\emph{keywords: Large-scale structure of Universe. Galaxies: clusters: general. Instabilities.}

\section{Introduction}
\noindent It is by now an established fact that the universe
accelerates faster than previously anticipated \cite{Krauss,Krauss2,Per1,Per2,Riess,Allen}. Hence some hitherto neglected
ingredient (in general called Dark Energy) has to be responsible for this new phenomenon. 
To account for this phenomenon we can introduce new physics in terms of a scalar field \cite{Ratra} or
modify three expressions in Einstein's equations which are often equivalent 
to a specific model with a scalar field. 
The three possibilities to account for the new physics are: the Einstein tensor, the energy-momentum
tensor (this is to say, the energy momentum tensor of a fluid 
gets modified by the inclusion of other components, \cite{Kremer} or the 
equation of state \cite{Kamenshchik}). 
The first possibility encompasses
a positive cosmological constant $\Lambda$ and higher order gravity
with a more complicated Einstein-Hilbert action \cite{Nojiri}. In the the present work we choose to work
with the cosmological constant as the simplest explanation for the acceleration of the universe.
We shall put forward the question if such a cosmological model has an influence 
on astrophysical structures. We shall use equilibria concepts like hydrostatic
equilibrium and virial equations to see how relatively low density astrophysical matter
of different shapes behaves in a fast expanding universe. Anticipating the results, we can say that
indeed there are some interesting effects.

\noindent Often it is assumed that $\Lambda$ does not have any effect on 
astrophysical processes which take place at scales different from the
cosmological ones.
Indeed, looking at the scales set by $\Lambda$, this assumption seems to be
justified at the first glance. The scales set by the cosmological constant
are of truly cosmological order of magnitude \cite{Nowakowski1,Bala1}). 
The density scale
is set by $\Lambda = 8\pi G_N \rho_{\rm vac}$ with $\rho_{\rm vac} \simeq (0.7-0.8)\rho_{\rm crit}$.
The length scale, $r_{\Lambda}=1/\sqrt{\Lambda}$ is of the order of the Hubble
radius while the mass scale $M_{\Lambda}=r_{\Lambda}/G_N$ reaches up to the value of the
mass of the universe. These scales constitute the so-called coincidence 
problem, namely the question as to why we should live exactly  at an epoch where
the scales of the cosmological constant are also the scales of the universe.
Neither was it so in the past nor will it be so in the future when the universe expands further.
The only astrophysical structures which match these scales are superclusters whose
densities
are indeed of the order of magnitude of $\rho_{\rm crit}$. Indeed, here we can almost be sure that
the cosmological constant is of relevance \cite{Now2,Sussman}.
However, probing into astrophysical consequences of the cosmological constant of other, smaller and denser
structures
like clusters of galaxies or even galaxies themselves, would look a hopeless undertaking
unless we find circumstances where the effect of $\Lambda$ (which in the very principle is present)
gets enhanced. This can indeed happen through various mechanisms.
For instance, in a problem where $r_{\Lambda}$ combines with a much smaller length scale,
say $r_0$,
the effects can be sometimes expressed as $r_0^nr_{\Lambda}^m$.  
In consequence, the observable in which this expression enters gets affected
by $\Lambda$ in a way which is important at much smaller scales 
than $r_{\Lambda}$. A concrete example is the Schwarzschild-de Sitter metric where
we find the parameter $r_{\Lambda}$ together with the much smaller length scale of the 
Schwarzschild radius $r_{s}$.
These two conspire in the form $(r_{s} r_{\Lambda})^{1/3}$ to define the largest extension of bound orbits as explained in the text and in \cite{Balaguera-Antolinez}.
We will discuss a very similar combination which emerges from the virial theorem defining
the largest possible virialized structure with a given mass.
Another possibility to enhance the effect of the
cosmological constant is to consider non-spherical objects. It then often happens
that the effect of $\Lambda$ becomes $(l_1/l_2)^k\rho_{\rm vac}$ where $l_i$
are two different length scales of a flattened object like a disk or an ellipsoid \cite{Now2}.
This  indeed happens for many astrophysical quantities, among other the critical mass,
the angular velocity and the temperature (mean velocity of the components
of the large structure) which we will discuss in the present paper. 
Finally, we can vary a dimensional variable
to see if this enhances the effect of the cosmological constant.
As an example let us quote the polytropic index $n$ in the equation of state
entering also the Lane-Emden equation. It is known that with growing $n$ ($n \ge 5$)
the object described by this equation of state does not have a well-defined radius
as the density goes only asymptotically to zero. We will show that this pattern of behaviour
becomes more dominant with $\Lambda \neq 0$.

\noindent Of course, all these effects become stronger the more diluted the mass
conglomeration is. The superclusters are certainly the best candidates if we look
for astrophysical effects of the cosmological constant. As a matter of fact,
they do not seem to be virialized due to the extreme low density and their
pancake structure \cite{Jaaniste} where the effect of flatness mentioned above becomes 
powerful \cite{Now2,Sussman}. For the next structure, the clusters of galaxies (or groups of galaxies)
with densities between one and three orders of magnitude above the critical density \cite{Padma1},
we would need one of the enhancing factors discussed above to see an appreciable effect of $\Lambda$.
This is possible in various ways as shown below. Galaxy clusters can have various forms, among others oblate and 
prolate \cite{Cooray, Kalinkov}.
And what is more, they can even rotate \cite{Kalinkov}. 
We will show an explicit effect of $\Lambda$ on their
angular velocity and temperature in case the angular velocity is zero. Since the effect
of large eccentricity is larger for prolate than for oblate ellipsoids, it is 
comforting to know that clusters can assume a prolate shape.
For low-density galaxies like the Low Surface Brightness (LSB) galaxies whose density
is roughly four orders of magnitude above the critical density \cite{Blok1,Blok2}, we still find some
effects. For $n=5$ and $\Lambda=0$ the solution of the Lane-Emden equation \cite{Chs} is very often used as a phenomenologically valid description of the density profile (called also 
Plummer's law \cite{Plu1}. This is still possible as $\rho \to 0$ as $r \to \infty$. However, this
property vanishes for low-density galaxies and the $n=5$ case not only does not have a well-defined radius,
its solution does not vanish asymptotically which thus rendering it unphysical.

\noindent The paper is organized as follows. In the second section we will briefly review
the general form of virial theorem including pressure, magnetic fields and, of course, the cosmological constant.
Here we will also discuss some general results regarding $\Lambda$. In the third section we will
specialize on spherical configurations. We will show how $\Lambda$ sets the scale of a maximal virial radius
and compare it to a result from the Schwarzschild-de Sitter metric. We will also solve the Lane-Emden
equation numerically and analytically (for the polytropic index $n=2$).  In the fourth section we will
discuss non-spherical configurations. First, we will show how $\Lambda$ affects the
angular velocity of spheroids. In addition, we will discuss the
effects of $\Lambda$ for the critical mass, mean velocity and mean rotational velocity too. 
The fifth  section is devoted to small oscillations around equilibrium.

\section{Local dynamics with cosmological constant}
\subsection{Newton-Hooke spacetime}
The cosmological constant enters the equations of Newtonian limit as a consequence of its appearance in the Einstein field equations.
It is through this weak field limit approximation that $\Lambda$ enters also in the equations describing the structure of astrophysical configurations. It is interesting to note that all variables to be found in the
virial equations, are also present in the Poisson equation of the Newtonian limit. However, this is not always
the reason why these terms enter the virial equations, at least in the first order. The Poisson equation for a self-gravitating system modeled as an ideal fluid is written as 
\begin{equation}
\label{poisson}
\nabla^{2}\Phi=4\pi G_{N}\lp \rho+3\frac{P}{c^{2}}+2\frac{\mau_{\rm em}}{c^{2}}\rp - \Lambda.
\end{equation}
where $P$ is the pressure and $\mau_{\rm em}$ is the electromagnetic energy density. 
The solution of (\ref{poisson}) at the zeroth order of $v/c$ (from now on we set $c=1$) is written as
\begin{equation}
\label{psolution}
\Phi(\vr)=-G_{N}\int_{V'}\frac{\rho(\vr')}{|\vr-\vr'|}\,\dd ^{3}r'-\frac{1}{6}\Lambda |\vr|^{2}+\cdots
\end{equation}
where the dots stand for the correction terms that appear because the boundary conditions are now set at a 
finite distance \cite{Nowakowski1}. These terms can be usually neglected. 
The cosmological constant $\Lambda >0$ contributes to the expansion of the universe. This fact remains partly valid in the Newtonian limit
where $\Lambda$ gives us an \emph{external force}. This defines the 
so-called (non-relativistic) Newton-Hooke spacetime \cite{Bacry,Derome, Aldrovandi}. \\
\subsection{The $\Lambda$-virial theorem}
The second order tensor virial equation can be derived in different ways: 
from a statistical point of view through the collisionless Boltzmann equation, from a variational principle or 
by direct differentiation of the moment of inertia tensor 
\begin{equation} \label{I}
\mai_{ik}=\int_{V}\rho(r) r_{i}r_{k}\,\dtr.
\end{equation}
In the following we use the statistical approach \cite{Cha1}
which also allows to derive higher order virial equations 
(for instance, the first order virial equation refers to the motion of the center of mass). In this context, 
from Boltzmann's equation one can derive the equation for momentum conservation (Euler's equation) written for 
a self gravitating system influenced by a magnetic field as 
\begin{equation}
\label{euler}
\rho \frac{\dd \langle v_{i}\rangle }{\dd t}+\rho \pai\Phi+\frac{1}{2}\pai(\magne^{2})+\paj\lp\mapp_{ij}-B_{i}B_{j}\rp=0,
\end{equation}
where $\Phi$ is the gravitational potential given by (\ref{psolution}) (which includes $\Lambda$) and 
\begin{equation}
\label{mapp}
\mapp_{ij}\equiv\rho\langle (v_{i}-\langle v_{i}\rangle )(v_{j}-\langle v_{j} \rangle)\rangle\
=\delta_{ik}P+\pi_{ik}, \hspace{0.8cm}{\rm Tr}(\pi_{ik})=0,
\end{equation}
is the pressure tensor, $P$ is the pressure and $\pi_{ik}$ its traceless part. Equations (\ref{poisson}) and (\ref{euler}) 
together with an equation of state $P=P(\rho,s)$ ($s$ is the entropy) complete the description of a self gravitating fluid. 
By taking exterior products of $r_{k}$ with Euler's equation and integrating over the volume of the system one obtains 
the second order virial equation as
\begin{equation}
\label{virotro}
\frac{1}{2}\frac{\dd ^{2}\mai_{ik}}{\dd t^{2}}=2T_{ik}-|\maw_{ik}^{\rm gen}|
+\frac{8}{3}\pi G_N \rho_{\rm vac}\mai_{ik}+ \Pi_{ik}\hspace{0.8cm}{\rm where} \hspace{0.3cm}\Pi_{ik}=\int_{V}\mapp_{ik} \dtr,
\end{equation} 
where $T_{ik}$ is the kinetic energy tensor and $|\maw_{ik}^{\rm gen}|$ 
is a generalized potential energy tensor which contains the contribution 
from the gravitational potential energy tensor $\maw_{ik}^{\rm N}$ 
and the  contributions of magnetic field through  
\begin{equation} 
\label{magB}
|\maw_{ik}^{\rm gen}|\equiv |\maw_{ik}^{\rm N}|(1-\Delta_{(ik)}).
\end{equation}
The other quantities are defined as follows
\begin{equation}
\label{mof}
T_{ik}\equiv \frac{1}{2}\int_{V}\rho \langle v_{i}\rangle \langle v_{k}\rangle \dtr, \hspace{0.2cm}
\maw_{ik}^{\rm N}=-G_{N}\int_{V}\rho(\vr)r_{i}\pak \Phi(\vr)\dtr, \hspace{0.2cm}\hspace{0.2cm}\Delta_{(ik)}
\equiv\frac{\maf_{ik}(\magne)}{|\maw_{ik}^{\rm N}|},
\end{equation}
together with 
\begin{equation}
\label{gama2}
\maf_{ik}(\magne)\equiv \delta_{ik}\mab-2\mab_{ik}-\int_{\pa V}r_{k}\left[\frac{1}{2}\delta_{ij}\magne^{2}-\Bi\Bj\right]\dd S_{j},\hspace{0.5cm}\mab_{ik}=\frac{1}{2}\int_{V}\Bi\Bk\,\dtr ,\hspace{0.5cm}\mab={\rm Tr}({\mab_{ik}}).
\end{equation}
A very useful version of the virial equation can be derived by assuming an isotropic pressure 
tensor and taking the trace in (\ref{virotro}). This way we get the scalar $\Lambda$-virial equation 
\begin{equation}
\label{scalar1}
\frac{1}{2}\frac{\dd^{2}\mai}{\dd t^{2}}=2\mak-|\maw^{\rm gen}|+\frac{8}{3}\pi G_N \rho_{\rm vac}\mai,
\end{equation} 
where the total kinetic energy is written as
\begin{equation}
\label{kin}
\mak=\frac{1}{2}\int_{V}\rho \langle v^{2}\rangle\, \dtr=T+\frac{3}{2}\Pi,\hspace{0.3cm}{\rm with}\hspace{0.3cm}\Pi\equiv \int_{V}P\,\dtr
\end{equation} 
The equilibrium condition is reached for $\ddot{\mai}=0$. This gives us the general $\Lambda$-virial theorem
\begin{equation}
\label{vt}
2T_{ik}-|\maw_{ik}^{\rm gen}|+\frac{8}{3}\pi\rho_{\rm vac}\mai_{ik}+
\Pi_{ij}=0,\hspace{0.8cm}2\mak-|\maw^{\rm gen}|+\frac{8}{3}\pi \rho_{\rm vac} \mai=0.
\end{equation} 
For rotating configurations with constant angular velocity, the kinetic 
term is modified as in the standard way as 
\begin{equation} 
\label{rotation}
T_{ik}\to T_{ik}+\mar_{ik},\hspace{0.5cm} \mar_{ik}\equiv \frac{1}{2}\lp\Omega_{\rm rot}^{2}\mai_{ik} -\Omega_{{\rm rot} i}
\mai_{kj}\Omega_{j}^{\rm rot}\rp,
\hspace{0.5cm} {\rm R}={\rm Tr}(\mar_{ij}),
\end{equation}
with ${\cal R}_{ij}$ the rotational kinetic energy tensor and $T_{ik}$ is referred to motions observed from the rotating reference frame. 
\noindent The $\Lambda$-virial theorem has been used in different 
contexts in \cite{Bala1,Now2,Barrow,Wang}. In the present work we will extend these studies.
\\
\subsection{General consequences}
The tensor virial equation is widely used in many astrophysical applications. 
The inclusion of $\Lambda$ provides a new way to study effects of the cosmological constant 
(parameters , in general) on astrophysical objects. The outcome depends essentially on two factors:
the geometry of the configuration and the density profile. 
We will explore the spherical geometry for both constant and varying density profiles 
and study some effects for non spherical geometry with constant density. 
The consequences that can be derived from the $\Lambda$-virial theorem can be classified 
in two categories. The first one puts an upper bound on the cosmological constant 
or alternatively a lower bound on density of objects in gravitational
equilibrium. Provided these bounds are satisfied, we can also study in the second step
the effects of $\Lambda$ on other properties of the astrophysical configurations like rotation, small oscillations etc.\\
The first simple consequence of the virial equation emerges if we require the system to satisfy (\ref{vt}). 
The fact that $\mak>0$ implies an upper bound on the vacuum energy density
\begin{equation}
\label{ineq0}
\rho_{\rm vac}\leq \frac{3}{8\pi\mai}\frac{|\maw^{\rm gen}|}{G_N}.
\end{equation} 
All systems in equilibrium have to satisfy (\ref{ineq0}).
Note that the right hand side  of this expression is a function of both the density and the geometry of the system. 
Hence, we must expect different bounds for different geometries and density profiles. 
For instance, if we assume a constant density and ${\bf B}=0$, we can define $\tilde{\maw}^{\rm N}$  and $\tilde{\mai}$ through 
\begin{equation} 
\label{tilde}
|\maw^{\rm N}|=\frac{1}{2}G_N \rho^{2}|\tilde{\maw}^{\rm N}|,\hspace{2cm}\mai=\rho\tilde{\mai},
\end{equation} 
such that the bound written in (\ref{ineq0}) becomes
\begin{equation}
\label{ineq}
\rho \geq \mathcal{A}\rho_{\rm vac}, \hspace{1cm}{\rm with} 
\hspace{1cm}\maa \equiv \frac{16\pi}{3}\lp\frac{\tilde{\mai}}{|\tilde{\maw}^{\rm N}|}\rp.
\end{equation} 
The factor $\maa$ which is only a function of the geometry (if we neglect the contribution of magnetic fields), 
will appear in many places in the paper. 
Its relevance lies in the fact that it enhances the effect of the cosmological constant for 
geometries far from spherical symmetry when ${\cal A}$ is large. A useful generalization 
can be done for situations in which we use the tensor form of the virial equation, namely
\begin{equation}
\label{agen}
\maa_{ij}\equiv \frac{16\pi}{3}\lp\frac{\tilde{\mai}_{ij}}{|\tilde{\maw}^{\rm gen}_{ij}|}\rp, \hspace{0.8cm}\rho={\rm constant}.
\end{equation} 
Finally, a curious equation can be derived by eliminating $\Lambda$ from the tensor virial equations:
\begin{equation} \label{curious}
\frac{2T_{ij} -|\maw^{\rm gen}_{ij}|+\Pi_{ij}}{2T_{nm} -|\maw^{\rm gen}_{mn}|+\Pi_{mn}}
=\frac{I_{ij}}{I_{mn}}.
\end{equation}
Although $\Lambda$ does not enter this equation, (\ref{curious}) is only valid if the denominator is non-zero as is the case
with $\Lambda \neq 0$. In Sect. 4 we will use this equality to infer a relation between the geometry and rotational velocity of 
an ellipsoid. Having discussed the general form of the virial theorem, we will discuss now the effects of $\Lambda$ and set $\magne=0$.

\section{Spherical configurations}
The tensor virial equation is trivially satisfied for spherically symmetric configuration without a magnetic field, 
since $\maw_{ik}^{N}=\delta_{ik}\maw^{N}$ and $\mai_{ik}=\delta_{ik}\mai$. Therefore, in this section we use only the scalar form of (\ref{virotro}).

\subsection{Constant density}
Explicit expressions can be derived in the spherical case with constant density, with  $|\maw^{\rm N}|=\frac{3}{5}\frac{G_{N}M^{2}}{R}$ and $\mai=\frac{3}{5}MR^{2}$, 
so that $\maa_{\rm spherical}=2$. In this case the ratio $\rho_{\rm vac}/\rho$ does not get enhanced much by the
geometrical factor $\maa$.

\noindent 
Worth mentioning is the result from general relativity.
There the upper bound for the cosmological constant comes out as $\Lambda \leq 4\pi G_{N}\bar{\rho}$ \cite{Bo2,Boehmer2003}
where $\bar{\rho}$ is the mean density defined by
$\bar{\rho}=M/V$.
This bound is derived not only from Newtonian astrophysics, 
but also from a general relativistic context via the Tolmann-Oppenheimer-Volkoff equation \cite{Oppenheimer}
for hydrostatic equilibrium of compact objects \cite{Bala1,Boehmer2003}.

\noindent Another relevant effect of the cosmological constant is the existence of a maximal virial radius of a
spherical configuration which can be
calculated 
from the $\Lambda$-virial equation. Using the expressions for $|\maw^{\rm N}|$ and $\mai$ given before, equation 
(\ref{scalar1}) yields as a cubic equation for the virial radius $R_{\rm vir}$ 
\begin{equation}
\label{cubic}
R_{\rm vir}^{3}+\lp10\eta \rl^{2}\rp R_{\rm vir}-3\rs \rl^{2}=0.
\end{equation} 
Here we introduced the dimensionless temperature $\eta$ as
\begin{equation}
\label{eta}
\eta\equiv \frac{\mak}{\rs}=\frac{3k_{B}}{\mu}T,
\end{equation} 
where $\mu$ is the mass of the average member of the configuration, $k_{B}$ is the Boltzmann constant, 
$T$ is the temperature and $r_s$ is defined by
\begin{equation} \label{rs}
r_s=G_NM,
\end{equation}
The length scale $\rl$ is set by the the cosmological constant as 
\begin{equation}
\label{rl}
\rl\equiv \sqrt{\frac{1}{\Lambda}}=2.4\times 10^{3}\, h_{70}^{-1}\, 
\Omega_{\rm vac}^{-1/2}\, \text{Mpc} \approx 1\times 10^{10}\, \, \text{ly}.
\end{equation} 
$h_{70}\approx0.7$ is dimensionless Hubble parameter \cite{Rich} and 
$\Omega_{\rm vac}=\rho_{\rm vac}/\rho_{\rm crit}\approx 0.7$ is the density parameter at the present time. 
The positive real root of equation (\ref{cubic}) is given by  
\begin{equation}
\label{rvir}
R_{\rm vir}(\eta)=\varpi (\eta) R_{\rm vir}(0),
\end{equation} 
where $R_{\rm vir}(0)$, is radius for the configuration at $\eta=0$ is given by
\begin{equation}
\label{rvir2}
R_{\rm vir}(0)=\lp 3 x\rp ^{1/3}\rl=(3r_s r_{\Lambda})^{1/3},
\end{equation} 
and the dimensionless parameter $x$ is defined as 
\begin{equation}
\label{x}
x=\frac{\rs}{\rl}=1.94\times 10^{-23}\lp\frac{M}{M_{\odot}}\rp 
 h_{70}\Omega_{\rm vac}^{1/2} \ll 1 .
\end{equation} 
The radius $R_{\rm vir}(0)$ is the largest radius that a spherical homogeneous cloud may have in virial equilibrium 
(i.e, satisfying (\ref{vt})). The function $\varpi(\eta)$ can be obtained from the solution of the cubic equation and reads
(clarified in the appendix A)
\begin{equation}
\label{delta}
\varpi(\eta)=2.53 x^{-1/3}\eta^{1/2}\sinh\left[\frac{1}{3}{\rm arcsinh}\lp0.24 x\eta^{-3/2}\rp\right].
\end{equation} 
Figure 1 shows the behavior of $\varpi(\eta)$ for different values of $x$. 
We see that the increase of the temperature implies a decrease of the effects of $\Lambda$
which can be easily checked if we solve $R_{\rm vir}$ from the virial theorem with $\Lambda=0$ and
compare it to the approximation $\eta\to \infty$ in (\ref{rvir}): 
\begin{equation}
\label{dapp}
R_{\rm vir}(\Lambda \to 0)=R_{\rm vir}(\eta \to \infty)\equiv R_{\rm vir}(\eta_{\star})=\frac{3}{10}\frac{\rs}{\eta_{\star}}.
\end{equation} 
We can consider (\ref{rvir}) as a  radius-temperature relation for a fixed mass applied on astrophysical structures in a single state of equilibrium
in the presence of $\Lambda$. That is, given $x$ and $\eta$ we calculate the radius. 
But we can adopt another point of view for this relation. 
Imagine a spherical configuration characterized by a constant mass $M$. In analogy to a thermodynamical reversible process, the configuration may pass from one state of virial equilibrium to another following the curve $\varpi-\eta$, that is, satisfying the condition $\ddot{\mai}=0$. Clearly, there must be some final temperature $\eta_{\star}$ when this process ends since the temperature cannot increase indefinitely. But of course since the virial equations are not dynamical we cannot know which stage is the final one. If we \emph{assume} that the effects of $\Lambda$ are negligible when $\eta=\eta_{\rm \star}$, then using Eqs (\ref{delta}) and (\ref{dapp}) we get
\begin{equation}
\label{temperature}
\eta_{\star}=0.208 \varpi_{\star}^{-1}x^{2/3},\hspace{0.8cm}\varpi_{\star}=\frac{R_{\rm vir}(\eta_{\star})}{R_{\rm vir}(0)}.
\end{equation} 
This is an equation for the temperature $T_{\star}$ 
\begin{equation}
\label{temperature2}
M^{-2/3}T_{\star}=0.138\lp\frac{\mu}{k_{B}}\rp\varpi_{\star}^{-1}\rl^{-2/3}.
\end{equation} 
For a hydrogen cloud ($\mu=m_{\rm proton}$), we then write the mass-temperature relation using Eqs.(\ref{rl}) and (\ref{x}) as
\begin{equation}
\label{temperature3}
T_{\star}=8.60\times 10^{-4}\varpi_{\star}^{-1}\lp \frac{M}{M_{\odot}}\rp^{2/3}h_{70}^{2/3}\Omega_{\rm vac}^{1/3}\hspace{0.2cm}{\rm K}.
\end{equation} 
Note that this expression maintains the same dependence of the standard mass-temperature relation derived 
from the virial theorem, i.e, $T\propto M^{2/3}$ (see \cite{Wang} or equation (\ref{mvel2}) of this paper). 
However the meaning of (\ref{temperature3}) is different from that of typical mass temperature relations since (\ref{temperature3}) 
is associated to the temperature that a system acquires in the final stage
after after going through some \emph{reversible processes}
which took the system through successive states of virial equilibrium 
with constant mass from a radius $R_{\rm vir}(0)$ to a radius $R_{\rm vir}(\eta_{\star})$ or vice versa. 
On the other hand, the 
mass-temperature relation like Eq. (\ref{mvel2}) of this paper relates the temperature of any configuration in equilibrium with its observed mass at constant density. In that context one considers only one equilibrium state and the cosmological constant enters just as corrections.

\begin{figure}
\begin{center}
\includegraphics[width=8cm,height=6.5cm]{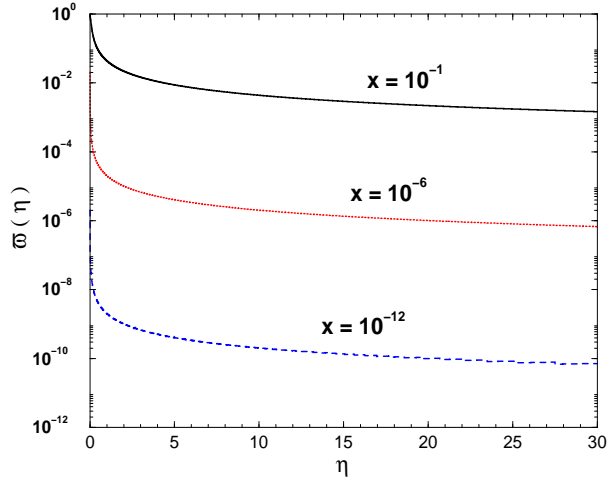}
\end{center}
\caption{\footnotesize{Ratio between $R_{\rm vir}(\eta)$ and $R_{\rm 
vir}(0)$ for different values of $x$ as a function of $\eta=(3/\mu)T$, where $\mu$ is the mass of the main average components of the system.}}\label{gdelta}
\end{figure}

\noindent As a final remark on Eq. (\ref{rvir}), we discuss a result which formally coincides with the 
virial radius $R_{\rm vir}(0)$ derived from the Schwarszchild de-Sitter 
spacetime \cite{Balaguera-Antolinez}, 
but whose physical meaning is quite different. 
The Schwarzschild-de Sitter metric takes the form 
\begin{equation}
\label{sca1}
\dd s^{2}=-\lp1-\frac{2R_{\rm s}}{r}-\frac{r^{2}}{3\rl ^{2}} \rp\dd t^{2}+\lp1-\frac{2R_{\rm s}}{r}-\frac{r^{2}}{3\rl ^{2}} \rp^{-1}\dd r^{2}+r^{2}\dd \theta^{2}+r^{2}\sin^{2}\theta \dd \phi^{2},
\end{equation}
Now in contrast to (\ref{rs}) we have 
\begin{equation}
R_s=G_{\rm N}\mu,
\end{equation}
with $\mu$ the mass of the object giving rise to the
Schwarzschild-de Sitter metric (in Eq. (\ref{rs}) $M$ is the mass of the total conglomeration whereas
here we consider $\mu$ as the mass of its average member).
Choosing the affine parameter as the proper time $\tau$ the equation of motion of a test-body
can be cast in a form similar to the corresponding equations from non-relativistic classical
mechanics. 
\begin{equation}
\label{sca2}
\frac{1}{2}\lp\frac{\dd r}{\dd \tau}\rp^{2}+U_{\text{eff}}=\frac{1}{2}\lp \mathcal{E}^{2}-1 \rp \equiv C=\text{constant},
\end{equation}
where $\mathcal{E}$ is a conserved quantity and $U_{\rm eff}$ is the effective potential, defined by
\begin{equation}
\label{sca2a}
\mathcal{E}=\lp1+2U_{\rm eff}(r)\rp\frac{\dd t}{\dd\tau},\hspace{0.5cm}U_{\text{eff}}(r)=-\frac{R_s}{r}-\frac{1}{6}\frac{r^{2}}{\rl ^{2}}.
\end{equation}
For simplicity we are have chosen here the angular momentum $L$ to be zero. With
$L$ zero or not, $U_{\rm eff}$ displays a local maximum below zero due
to $\Lambda \neq 0$ forming a potential barrier. This is to say, the standard local minimum where we find all the bounded orbits is now followed by a local maximum after which $U_{\rm eff}$ goes to $-\infty$. With $\Lambda=0$ this function approaches zero asymptotically.
One is immediately tempted to say that this barrier will occur at
cosmological distances. This is not the case and one calculates
\begin{equation}
\label{sca3}
r_{\text{max}}=\lp 3R_s \rl ^{2}\rp^{1/3}=
9.5\times 10^{-5}\lp\frac{\mu}{M_{\odot}}\rp^{1/3}\lp
\frac{\rho_{\text{crit}}}{\rho_{\text{vac}}}\rp^{1/3} h_{70}^{-2/3} 
\, \text{Mpc}.
\end{equation}
In other words, the combination of the large scale $r_{\Lambda}$ with the small
scale $R_s$ gives us a distance of astrophysical relevance, namely $r_{\rm max}$. Its relevance lies in te fact that beyond $r_{\rm max}$ there are no bound orbits. Indeed, with
$\mu$ the solar mass, $r_{\rm max}$ is of the order of a globular cluster extension ($70$ pc); 
with $\mu$ as the mass of globular cluster, $r_{\rm max}$ comes out to be of the order
of the size of a galaxy ($10$ kpc), and finally taking $\mu$ to be the mass of a galaxy, $r_{\rm max}$
gives the right length scale of a galaxy cluster ($1$ Mpc). 
Certainly, the value of the extension of a large astrophysical body is the result of a multi-body interaction. But with the actual values of $r_{\rm max}$, it appears as if the \emph{length scale} (we emphasize that we are concerned hare about scales and not precise numbers) of an astrophysical conglomeration is approximately $r_{\rm max}$, which apparently means that this scale does not change drastrically when going from a two body problem to a multi-body calculation.  This makes sense if the object under consideration is not too dense. We can now say that whereas $M$ in $R_{\rm vir}(0)$ (via $r_s$) is the mass of the object,
$\mu$ in Eq. (\ref{sca2}) is the mass of its members. Clearly, we have $R_{\rm vir}(0) \ll r_{\rm max}$,
but both scales are of astrophysical order of magnitude. A result related to (\ref{rvir2}) derived in the framework of general 
relativity can be found in \cite{Mak}.
\subsection{Non-constant density}
The examination of configurations with non-constant densities can be done in two directions. 
Knowing the density profile $\rho(r)$, we can set up the virial equation and evaluate 
the equilibrium conditions from the inequality (\ref{ineq0}). 
In this picture, the effects of $\Lambda$ are included in the solution for the potential $\Phi$ as in Eq. (\ref{psolution}) 
and the resulting term acts like an external force, as mentioned  before.\\
The second option is to combine the Eqs. (\ref{poisson}), (\ref{euler}) and an equation of state (e.o.s) 
for which we can take a polytropic form $P=\kappa\rho^{1+\frac{1}{n}}$. This way we obtain the Lane-Emden equation
with $\Lambda$ \cite{Bala1}
\begin{equation}
\label{le}
\frac{1}{\xi^{2}}\frac{\dd }{\dd \xi}\lp \xi^{2}\frac{\dd \psi}{\dd \xi}\rp+\psi^{n}=\zeta_{\rm c}
\hspace{0.8cm}
\zeta_{\rm c} \equiv 2\lp\frac{\rho_{\rm vac}}{\rho_{\rm c}}\rp,
\end{equation} 
where $\rho_{c}$ is the central density, $r= a \xi$, $\rho(r)=\rho_{\rm c} \psi^{n}(\xi )$ with $\psi(0)=1$, $\psi'(0)=0$ 
and $a$ is the associated Jeans length defined as
\begin{equation}
\label{a}
a\equiv \sqrt{\frac{\kappa (n+1)}{4\pi \rho_{\rm c}^{1-\frac{1}{n}}} }.
\end{equation} 
It is important to notice that in this picture the expected effects of $\Lambda$ are to be found in the behavior of the density profile 
since now Eq. (\ref{le}) implies that its solution is also  function of the parameter $\zeta_{\rm c}$. 
Some effects of $\Lambda$ are contained in the total mass and the radius of the configuration
which is reached when $\psi(\xi_{1})=0$. Hence (\ref{le}) implies 
\begin{equation}
\label{radius}
\xi_{1}=\left[\frac{1}{\zeta_{\rm c}}\left|\frac{\dd}{\dd \xi}\lp\xi^{2}\frac{\dd \psi}{\dd \xi} \rp_{\xi_{1}} \right| \right]^{1/2}
,\hspace{0.5cm}R=a\xi_{1}=\rl \left[\kappa(n+1)\rho_{\rm c}^{\frac{1}{n}} 
\left|\frac{\dd}{\dd \xi}\lp\xi^{2}\frac{\dd \psi}{\dd \xi} \rp_{\xi_{1}} \right|\right]^{1/2}.
\end{equation}
Note that the radius of the configuration is now proportional to $\rl$. This is due to the fact that $\Lambda$ sets a scale for length. 
However, this does not mean that $R$ will be always of the order $r_{\Lambda}$ as $\Lambda$ is also contained in 
the expression in the square brackets in Eq. (\ref{radius}). We will show this below in a concrete example.
Since $\Lambda$ is a new constant scale the Lane-Emden equation loses some of its scaling properties as explained
in \cite{Bala1}. 
The mass of the configuration can be determined as usual with,
\begin{equation}
\label{mass}
M(\xi)=4\pi a^{3}\rho_{\rm c}\int_{0}^{\xi}\xi^{2}\psi^{n}\,\dd \xi=
\frac{4}{3}\pi a^{3}\xi^{3}\rho_{\rm c}\left[\zeta_{\rm c}-\frac{3}{\xi}\frac{\dd \psi}{\dd \xi}\right],
\end{equation}
where we used Eq. (\ref{le}) for the second equality. 
The total mass is then obtained by evaluating the last expression at $\xi=\xi_{1}$. 
As expected, the mass increases because the Newtonian gravity has to be stronger in order for the configuration to be in equilibrium 
with $\Lambda\neq0$. Figure 2 shows the numerical solutions
for $n=1$ to $n=5$. We expect that the radius of the configuration is increased by the contribution of $\zeta_{\rm c}$
and find it confirmed in the figures. 
However, not always is the radius of the configuration well defined, even if $n<5$. 
For sizeable values of $\zeta_{\rm c}$ (black line) we cannot find physical solutions of Eq. (\ref{le}) 
as the function $\psi$ acquires a positive slope. 
One might be tempted to claim that the radius of the configuration could be defined in these situations as the position 
where $\psi$ has its first minimum, but as can be seen for $n=3$
such a radius would be smaller than the radius with $\zeta_{\rm c}\to 0$ 
which contradicts the behaviour shown for the other solutions where $\xi(\zeta_{\rm c}\neq 0)>\xi(\zeta_{\rm c}=0)$. 
As already mentioned above this is the correct hierarchy between the radii because large $\zeta_c$ gives rise to
a large external force pulling at the matter.
The numerical solutions show that for relatively large values of
$\zeta_{\rm c}$, only $n=1$ has a well defined radius. 
In this case the
effect of $\Lambda$ is a $13\%$ increase of the matter extension as compared to $\zeta_{\rm c}=0$. 
As we increase the polytropic index, $\zeta_{\rm c}\approx 10^{-1}$ leads to non-physical solutions 
while the effect with bigger values of $\zeta_c$ becomes visible only for $n=3$.
For instance,  $\zeta_{\rm c}\approx 10^{-3}$ results in a radius which is $17\%$ bigger than the corresponding value with $\zeta_{\rm c}\to 0$. 
The combination $n=4$ and $\zeta_{\rm c}\approx 10^{-3}$ also leads to a non-physical solution. 
whereas the radius of the case
$\zeta_{\rm c}\approx 10^{-4}$ displays a difference of $13\%$ as compared to $\zeta_{\rm c}\to 0$. 
Finally, for $n=5$, the only physical solutions are obtained for the lowest values of $\zeta_{\rm c}$ where $\psi'<0$
This case is particularly interesting as with $\Lambda =0$ it is often used a a viable phenomenological
parametrization of densities \cite{Plu1,Binney}.  The solution has an asymptotic behaviour as $r^{-a}$
which has been also found in LSB galaxies de Blok at al. 2004). With $\Lambda \neq 0$ the $n=5$ seems less appealing as the matter
is diluted. 
For all values of $n$, the difference between $\zeta_{\rm c}=10^{-5}$ and $10^{-7}$ is negligible.\\
\begin{figure}
\begin{center}
\label{lane}
\includegraphics[angle=270,width=14cm]{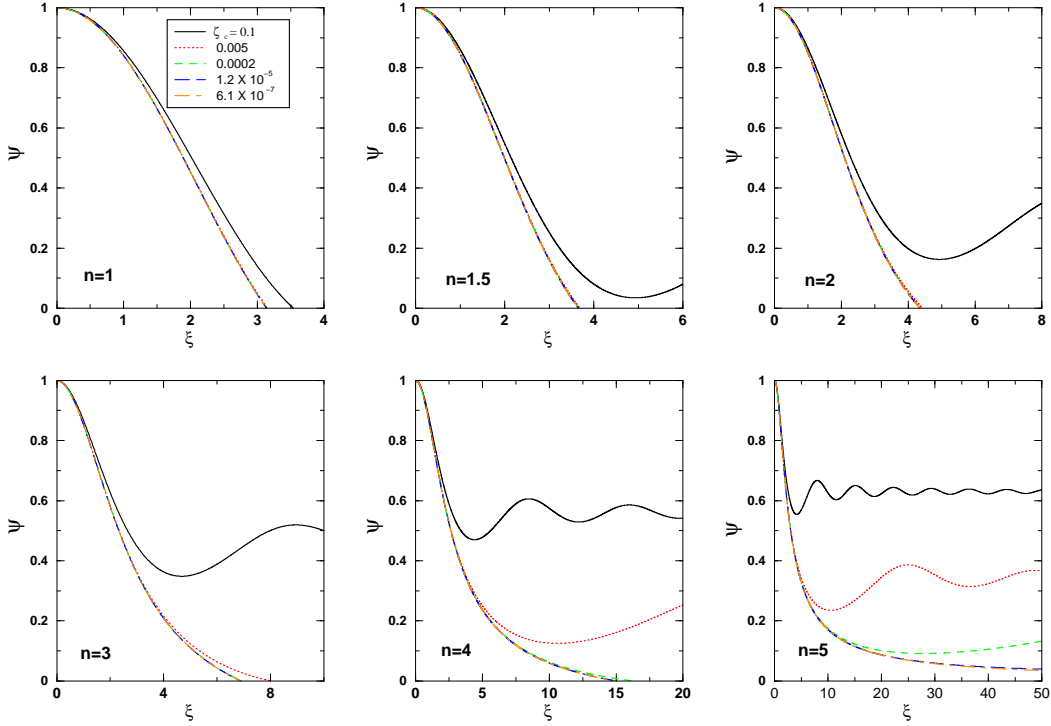}
\end{center}
\caption{\footnotesize{Effects of $\Lambda$ on the behaviour of the density of a polytropic configuration for different ratios $\zeta_{\rm c}$ 
and different polytropic indices. The radius of the configuration is not always defined, even for $n<5$. 
For higher values of $\rho_{\rm vac}$, only the $n=1$ case has a definite radius for these values of $\zeta_{\rm c}$. 
For other cases, the configuration is defined only for small $\zeta_{\rm c}$.}}
\end{figure}

\noindent Analytical solutions of (\ref{le}) can be found for $n=0,1$ and $n=5$ if $\Lambda=0$. For $n\to \infty$, 
the polytropic equation of state reduces to the equation of state 
of the isothermal sphere $P=\kappa \rho$. As an example, for $\Lambda\neq 0$, we can write the analytical solution in the case $n=1$ as
\begin{equation}
\label{n1}
\psi(\xi)=(1-\zeta_{\rm c})\frac{\sin \xi}{\xi}+\zeta_{\rm c}.
\end{equation}
The radius is $R=a\xi_{1}$, where $\xi_{1}$ is the solution of the transcendental equation  
\begin{equation} \label{rad1} 
\xi_{1}\zeta_{\rm c} =-(1-\zeta_{\rm c})\sin \xi_{1}.
\end{equation} 
In the first order of $\zeta_{\rm c}$ one finds 
\begin{equation}
\label{rn1}
R=\sqrt{\frac{1}{2}\pi\kappa}\lp1+\zeta_{\rm c}\rp.
\end{equation}
Equation (\ref{rad1})  also implies that there exists some $\zeta_{\rm 
crit}$ such that for $\zeta_{\rm c}\geq \zeta_{\rm crit}$, 
we cannot find a real solution for $\xi_{1}$. Approximately this gives
\begin{equation} \label{boundc} 
\rho_{c}\geq 10.8 \rho_{\rm vac},
\end{equation}
which, provided the overall density is not too big, is better that $\rho \ge 2\rho_{\rm vac}$ which is a result from
the general inequality (\ref{ineq}) for $\rho={\rm const}$ and spherical symmetry.
\noindent Finally, we can calculate the contribution of $\Lambda$ to the total energy of the object. Generalizing
the results found in \cite{Chs} we obtain
\begin{equation} 
\label{energy}
E=\frac{1}{3}(3-n)\maw^{\rm N}, \hspace{0.5cm} \maw^{\rm N}=-G_N\frac{M^2}{R}\left(\frac{3}{5-n}\right)
\left[1
+\left(\frac{R}{r_{\rm max}}\right)^3\right].
\end{equation}
from which we infer that the correction is very small in this case.\\
At the end of this section  we would like to summarize the findings from Fig. 2. In table 1 we write the ratio $\xi_{1}(\Lambda=0)/\xi_{1}(\Lambda\neq 0)$ for the same ratios $\zeta_{\rm c}$ and polytropic index as in Fig. 2. The horizontal line represents a non-defined radius. The symbol $\infty$ indicates that the radius is defined only asymptotically in case of $\Lambda=0$
\begin{table}
\begin{center}
\begin{tabular}{ccccccc}\hline \hline
$\zeta_{c}$         &$n=1$       &$n=3/2$     &$n=2$         & $n=3$         & $n=4$        & $n=5$ \\ \hline
$0.1$               &$\sim 0.88$ &--          &--            &  --           & --           &--    \\
$0.005$             &$\sim 1$    &$\sim 1$    &$\sim 0.98$   & $\sim 0.86$   & --           &--   \\
$2\times 10^{-4}$   &$\sim 1$    &$\sim 1$    &$\sim 1$      & $\sim 1$      & $\sim 0.88$  &--        \\
$1.2\times 10^{-5}$ & $\sim 1$   &$\sim 1$    &$\sim 1$      & $\sim 1$      & $\sim 1$     &--          \\ 
$6.1\times 10^{-7}$ & $\sim 1$   &$\sim 1$    &$\sim 1$      & $\sim 1$      & $\sim 1$     & $\infty$         \\
\hline
\end{tabular}
\end{center}
\caption[Fraction $\xi_{1}(\Lambda=0)/\xi_{1}(\Lambda \neq 0)$ for different $\zeta_{\rm c}$ polytropic index.]{\footnotesize{Values of the fraction $\xi_{1}(\Lambda=0)/\xi_{1}(\Lambda \neq 0)$ for different values of the ratio $\zeta_{\rm c}$ and the polytropic index. The horizontal lines represents the non well defined radius.}}\label{tablelane}
\end{table}

\section{Nonspherical configurations}
\subsection{Rotating configurations}
\begin{figure}
\begin{center}
\label{ag}
\includegraphics[angle=270,width=10cm]{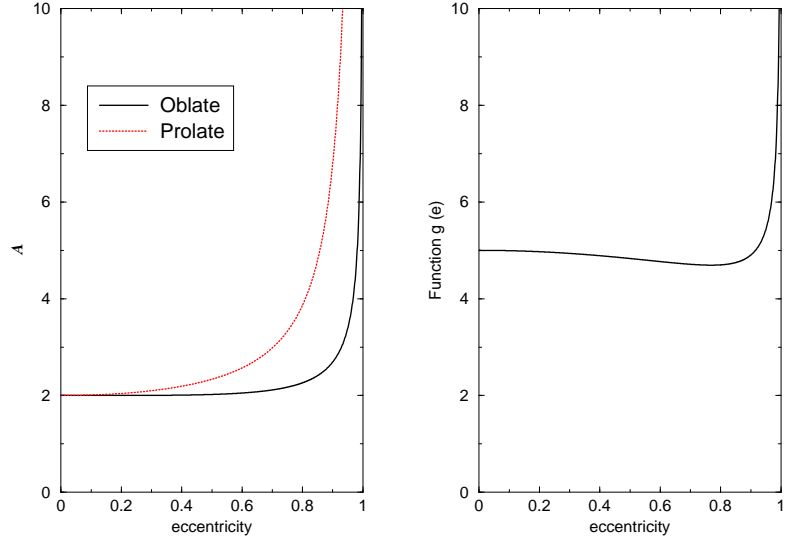}
\end{center}
\caption{\footnotesize{Function $g(e)$ and geometrical factor $\maa$(e) for prolate and oblate ellipsoids. 
These functions have their largest values for flat oblates and \emph{large} prolates.}}
\end{figure}

As emphasized already before, the effect of $\Lambda$ can get enhanced for non-spherical objects. This happens when the vacuum energy gets multiplied
by a ratio of two length scales $l_1$ and $l_2$ and we end up with expressions like $\rho_{\rm vac}(l_{1}/l_{2})^n$. 
For instance, for ellipsoidal configurations the geometrical parameter ${\cal A}$ entering among others the inequality (\ref{ineq}) can be calculated from its definition (\ref{agen}) with $\tilde{\mai}_{ij}=\frac{4}{15}\pi a_{1}a_{2}a_{3}\delta _{ij}a_{i}^{3}$ and $\maw_{ik}$ given in \cite{Binney}. We have \cite{Now2}
\begin{eqnarray} \label{A}
\maa_{\rm obl}&=&\frac{4}{3}\lp\frac{3-e^{2}}{\arcsin e}\rp\frac{e}{2\sqrt{1-e^{2}}} 
\stackrel{ a_{1} \gg a_{3}}{\to}  \frac{8}{3\pi}\lp\frac{a_{1}}{a_{3}}\rp, \nonumber \\
\maa_{\rm pro}&=&
\frac{4}{3}\frac{e(3-2e^{2})}{(1-e^{2})^{3/2}}\left[\ln\lp\frac{1+e}{1-e}\rp\right]^{-1} 
 \stackrel{ a_{3} \gg a_{1}    }{\to} \frac{2}{3}\lp\frac{a_{3}}{a_{1}}\rp^{3}\left[\ln\lp\frac{2a_{3}}{a_{1}}\rp\right]^{-1},
\end{eqnarray}
where the eccentricity\footnote{Once the density is given, the inequality (\ref{ineq}) becomes in this way a
defining equality for $e_{\rm max}$ such that $e <e_{\rm max}$ in order to maintain equilibrium.} 
$e$ is $e^{2}=1-a_{3}^{2}/a_{1}^{2}$ for the oblate and
$e^{2}=1-a_{1}^{2}/a_{3}^{2}$ for the prolate case. The behaviour of $\maa_{\rm obl}$ and $\maa_{\rm pro}$ is shown in figure \ref{ag}. 
It is clear that for very flat astrophysical objects we can gain in this way several orders of magnitude of enhancement 
of the effect of $\Lambda$ if ${\cal A}$
is a factor of attached to $\rho{\rm vac}$. 
Needless to say that we often encounter in the universe flat objects like elliptical galaxies, 
spiral disk galaxies, clusters of galaxies of different forms and finally 
superclusters which can have the forms
of pancakes.
Of course, the more dilute the system is, the bigger the effect of $\Lambda$. We can expect sizeable effects for clusters and superclusters,
even  for very flat galaxies. 
Regarding the latter, low density galaxies like the nearly invisible galaxies are among other the best candidates.

\noindent A convenient way to model almost all flat shaped objects is to consider ellipsoids which in the limit of
flattened spheroids can be considered as disks.  
There are three different kinds of elliptical configurations, characterized by three semi-axes 
$a_1=a$, $a_2=b$ and $a_3=c$: oblate , with $a=b<c$, prolate with $a=b>c$ and triaxial systems with $a> b > c$. 
Here the tensor virial equation provides a tool to determine which of these geometries are compatible with the  
virial equilibrium. Considering the case $\Lambda =0$ and $\rho=$constant, 
or spheroids with confocal density
distribution whose isodensity surfaces are similar concentric ellipsoids \cite{Binney,Roberts} i.e.
\begin{equation} \label{density1}
\rho=\rho(m^2), \,\,\, m^2=a_1^2\sum_1^3\frac{x_i^2}{a_i^2},
\end{equation}
the oblate ellipsoid (MacLaurin) emerges as a solution of the virial equations with a bifurcation point to a triaxial
Jacobi ellipsoid \cite{Che}. We can view the virial equation without $\Lambda$ as a homogeneous 
equation. Switching on $\Lambda \neq 0$ this
becomes an inhomogeneous equation whose right hand side is proportional to $\rho_{\rm vac}$. It is therefore a priori not clear
if in the case of $\Lambda \neq 0$ we can draw the same conclusions as with $\Lambda=0$.  \\
\noindent Let us assume a configuration which is rotating with constant angular velocity around the $z$ axis. 
Neglecting the internal motions,  the $\Lambda$-tensor virial Eq. (\ref{virotro}) for such a configuration is
\begin{equation}
\label{ns0}
\Omega_{\rm rot}^{2}\lp\mai_{ik}-\delta_{iz}\mai_{zk}\rp-|\maw_{ik}^{\rm N}|+\frac{8}{3}\pi G_N \rho_{\rm vac}\mai_{ik}
=-\delta_{ik}\Pi.
\end{equation}
The equations with diagonal elements yield the following identities
\begin{equation}
\label{ns01}
\Omega_{\rm rot}^{2}\mai_{xx}-|\maw_{xx}^{\rm N}|+\frac{8}{3}\pi G_N\rho_{\rm vac}\mai_{xx}=\Omega_{\rm rot}^{2}\mai_{yy}
-|\maw_{yy}|+\frac{8}{3}\pi G_N\rho_{\rm vac}\mai_{yy}=-|\maw_{zz}|+\frac{8}{3}\pi \rho_{\rm vac}\mai_{zz}.
\end{equation}
This set of equations can be resolved for the angular velocity:
\begin{equation}
\label{ns01}
\Omega_{\rm rot}^{2}=\lp\frac{|\maw_{xx}^{\rm N}|-|\maw_{zz}^{\rm N}|}{\mai_{xx}}\rp-
\frac{8}{3}\pi G_N\rho_{\rm vac}\lp1 -\frac{\mai_{zz}}{\mai_{xx}}\rp.
\end{equation}
The same expression holds if we make the replacement $\maw_{xx}\to\maw_{yy}$ and $\mai_{xx}\to\mai_{yy}$ 
on the right hand side of (\ref{ns01}). 
We can also eliminate $\Omega_{\rm rot}^{2}$ from these expressions to arrive at the condition 
\begin{equation}
\label{ns01a}
\delta_I\left[\frac{|\maw_{xx}^{\rm N}|-|\maw_{zz}^{\rm N}|}{|\maw_{yy}^{\rm N}|-|\maw_{zz}^{\rm N}|}\right]-1
=\frac{8\pi}{3}G_N\rho_{\rm vac}\mai_{zz}\lp\frac{1-\delta_I}{|\maw_{yy}^{\rm N}|-|\maw_{zz}^{\rm N}|}\rp.
\end{equation}
with $\delta_I\equiv \mai_{yy}/\mai_{xx}$. This expression  determines the possible ellipsoidal configurations. One solution is obvious: $\delta_{I}=1$ together with
$|\maw_{xx}^{\rm N}|=|\maw_{yy}^{\rm N}|$. This happens if the density is constant or for a 
confocal ellipsoid with the density given in (\ref{density1}). Hence this is not different from the
$\Lambda=0$ case. However, the bifurcation point to a triaxial system will get affected by $\Lambda$ \cite{Balab}.
Furthermore, the explicit calculation of this bifurcation point will now depend also on the details
of the density profile even if we take the latter to be as in Eq. (\ref{density1}). This is a direct consequence
of a `pre-existing' density scale $\rho_{\rm vac}$. For a homogeneous configuration we have
\begin{eqnarray}
\label{ns02}
\frac{\Omega_{\rm rot}^{2}}{8\pi G_N\rho}&=&\frac{1}{3}\maa_{xx}^{-1}\lp1-\eta_{x}\rp
\left[1-\lp\frac{\rho_{\rm vac}}{\rho}\rp\lp\frac{\maa_{xx}-\eta_{x}\maa_{zz}}{1-\eta_{x}}\rp\right]\\ \nonumber
\tilde{\delta}_I\lp\frac{\eta_{y}}{\eta_{x}}\rp\lp\frac{1-\eta_{x}}{1-\eta_{y}}\rp-1&=&
\lp\frac{\rho_{\rm vac}}{\rho}\rp \maa_{zz}\eta_{y} \lp \frac{1-\tilde{\delta}_I}{1-\eta_{y}}\rp.
\end{eqnarray}
where $\tilde{\delta}_I\equiv \tilde{\mai}_{yy}/\tilde{\mai}_{xx}$, 
$\eta_{i}=|\tilde{\maw}_{zz}^N|/|\tilde{\maw}_{ii}^N|$ and ${\cal A}_{ii}$ are defined in Eq. 
(\ref{agen}). 
For an oblate configuration they can be calculated in terms of the eccentricity $e=\sqrt{1-a^2/c^2}$ to give \cite{Binney}
\begin{equation}
\label{quant}
\maa_{xx}=\maa_{yy}=\frac{4}{3}\frac{e^{2}}{\sqrt{1-e^{2}}}\left[\frac{\arcsin e}{e}-\sqrt{1-e^{2}}\right]^{-1},\hspace{0.8cm}
\maa_{zz}=\frac{8}{3}\frac{e^{2}}{\sqrt{1-e^{2}}}\left[\frac{1}{\sqrt{1-e^{2}}}-\frac{\arcsin e}{e}\right]^{-1}.
\end{equation}
Furthermore we have
\begin{equation}
\label{eta2}
\eta_{x}
=\eta_{y}=2(1-e^{2})^{1/2}\lp\frac{e-\sqrt{1-e^{2}}\arcsin e}{\arcsin e-e\sqrt{1-e^{2}}}\rp.
\end{equation}
Using these expressions, we write the constant angular velocity for an oblate ellipsoid from the first line of equation Eq. (\ref{ns02}) as
\begin{equation}
\label{av}
\Omega_{\rm rot}^{2}=\Omega_{0}^{2}\left[1-\lp\frac{\rho_{\rm vac}}{\rho}\rp g(e)\right],
\end{equation}
where $\Omega_{0}$ corresponding to the angular velocity when $\Lambda=0$ is given by the Maclaurin formula
\begin{equation}
\label{omcero}
\Omega_{0}^{2}=\frac{8}{3}\pi\maa_{xx}^{-1}\lp1-\eta_{x}\rp=2\pi G_{N}\rho 
\left[\frac{(1-e^{2})^{1/2}}{e^{3}}(3-2e^{2})\arcsin e-\frac{3}{e^{2}}(1-e^{2})\right]
\end{equation}
The function $g(e)$ defined through Eq. (\ref{ns02}) can also be calculated with explicit 
dependence on the eccentricity as
\begin{equation}
\label{H}
g(e)\equiv \frac{\maa_{xx}-\eta_{x}\maa_{zz}}{1-\eta_{x}}=\frac{4}{3}e^{5}\left[(1-e^{2})^{1/2}(3-2e^{2})\arcsin e-3e(1-e^{2})\right]^{-1}.
\end{equation}
As is evident from the above equations, $\Lambda$ has a twofold effect on the angular velocity.
Firstly, it reduces the angular velocity with respect to the value $\Omega_{0}$ especially at the local maximum (see Figure 3).
This is not a small effect and can affect even galaxies. 
Secondly, we see from Eq. (\ref{omcero}) that $\Omega_0 \to 0$ for $e \to 1$. On the other hand 
\begin{equation}
\label{om}
\frac{\rho_{\rm vac}}{\rho}g(e) \to \frac{\rho_{\rm vac}}{\rho}\frac{32\pi}{9}\left(\frac{a_{3}}{a_{1}}\right),
\end{equation}
approaching $1$ for a very flat oblate configuration and not too dense matter. Therefore, beyond the local maximum in $\Omega_{\rm rot}$ 
the cosmological constant causes a  steeper fall of $\Omega_{\rm rot}$ towards $0$.
\noindent Another relevant interesting quantity which can be calculated in this context is the ratio of the rotational over 
the gravitational energy contributions to the scalar virial equations, i.e., $\mar/|\tilde{\maw}^N|$. 
In accordance with Eq. (\ref{omcero}) the latter can be written as
\begin{equation}
\label{beta}
\beta\equiv \mar/|\maw^N|, \hspace{0.8cm}
\beta=\beta_{0}\left[1-\lp\frac{\rho_{\rm vac}}{\rho}\rp g(e)\right],
\hspace{0.8cm}\beta_{0}=\frac{3}{2e^{2}}\left[1-\frac{e(1-e^{2})^{1/2}}{\arcsin e}\right]-1.
\end{equation}
The effects on $\beta$ are therefore similar to the the ones encountered in $\Omega_{\rm rot}$ (see also Figures 3). 
Finally, on account of $P >0$ we can infer from the virial equations with $\rho=$const the following inequality
\begin{equation}
\label{betaineq}
0\leq\beta\leq \frac{1}{2}-\lp\frac{\rho_{\rm vac}}{\rho}\rp\maa,
\end{equation}
which together with (\ref{beta}) results in an inequality for the density of an ellipsoidal configuration:
\begin{equation}
\label{bineq}
\rho\geq \tilde{\maa}\rho_{\rm vac},\hspace{0.8cm}\tilde{\maa}\equiv \frac{1}{2}\lp \frac{\maa-4\beta_{0}g}{1-2\beta_{0}}\rp.
\end{equation}
For $e\to0$, we have $\maa\to 2\tilde{\maa}$, while for $e\to 1$, $\maa\to \tilde{\maa}$, and $\maa \geq \tilde{\maa}$. 
Therefore the above inequality is slightly weaker than the bound given in (\ref{ineq}). Nevertheless it is useful since it it derived
directly from a different starting point ($P>0$) than (\ref{ineq}) which is based on ${\cal K} >0$.
\subsubsection{Mean mass-weighted rotational velocity}
Not always the deviation from spherical symmetry guarantees that the effect of $\Lambda$ on observables is
sizeable. This depends on the context and also which scales we compare. If we compare $\rho_{\rm vac}$ to densities,
the quantities ${\cal A}$ and ${\cal A}_{ii}$ defined in Eq. (\ref{ineq}) 
are for flattened objects large enough to enhance the effect of $\Lambda$. If $r_{\Lambda}$ is combined with the
Schwarzschild radius $r_s$ to give $r_{\rm max}$ as in Eq. (\ref{rvir2}), the result is still of astrophysical relevance.
But if we had to compare one of the axes of an ellipsoid to $r_{\Lambda}$ (i.e. $a_i/r_{\Lambda}$) the effect would be 
negligible unless the extension $a_i$ is bigger than Mpc (clusters and superclusters) and the small ratio
$a_i/r_{\Lambda}$ is comparable to another small quantity of the same order of magnitude entering the equation under
consideration.
This happens for instance if we generalize a result (discussed \cite{Binney}) 
on a mass-weighted mean-square rotation speed $v_0$ of an ellipsoidal object to include $\Lambda$. Assume that due to symmetry properties of the object
the only relevant components of the tensors in the tensor virial equations are $xx$ and $zz$. We then
obtain 
\begin{equation} 
\label{s1}
\frac{2T_{xx} + \Pi_{xx} +\frac{1}{3}\Lambda \mai_{xx}}{2T_{zz} + \Pi_{zz} +\frac{1}{3}\Lambda \mai_{zz}}
=\frac{|\maw_{xx}^{\rm N}|}{|\maw^{\rm N}_{zz}|}.
\end{equation}
If the only motion is a rotation about the $z$ axis we have $T_{zz}=0$ and we can solve for $T_{xx}$ as
\begin{equation} \label{txx}
2T_{xx}=\frac{|\maw_{xx}^{\rm N}|}{|\maw_{zz}^{\rm N}|}\left(\Pi_{zz} +\frac{1}{3}\Lambda \mai_{zz}\right) -
\left(\Pi_{xx} +\frac{1}{3}\Lambda \mai_{xx}\right).
\end{equation}
Using Eqs. (\ref{mapp}), (\ref{virotro}) and (\ref{mof}), the quantities entering our equation can be parametrized 
in the following way
\begin{equation} 
\label{para}
2T_{xx}=\frac{1}{2}\int\rho \langle v_{\phi}^{2}\rangle \dtr=\frac{1}{2}Mv_0^2,\hspace{0.5cm}\Pi_{xx} = M\sigma_{0}^{2},\hspace{0.5cm}\Pi_{zz} =(1-\delta_0)\Pi_{xx}.
\end{equation}
where $v_{0}^{2}$ is the mass weighted mean angular velocity, $\sigma_0$ is the mass-weighted mean-square random velocity in the $x$ direction and $\delta_0$ measures the anisotropy in $\Pi_{ii}$. 
If $\delta_0$ is of the order of one, it suffices to compare $\sigma_0^2$ with $\Lambda \mai_{xx}/3M$. If both are
of the same order of magnitude, the effect of $\Lambda$ is non-negligible. Since $v_0$ and $\sigma_0$ are of
the same order of magnitude and $v_0$ is non-relativistic, we can assume that $\sigma_0 < 10^{-2}$.  The
quantity $\Lambda I_{xx}/3M$ can be estimated to be $(a_1/r_{\Lambda})^2$. Hence if $a_i \sim 1 {\rm Mpc}$,
$v_0$ ($\sigma_0$) has to be truly non-relativistic and of the the order of $10^{-6}$ to gain an appreciable effect of $\Lambda$. This
improves if $a_i$ is one order of magnitude bigger which is possible for large galaxy clusters. The velocities have to be then at most
$10^{-4}$. In these cases we have to keep $\Lambda$ and while solving (\ref{para}) for $v_0^2$ one has
\begin{equation} 
\label{v0}
\frac{1}{2}\frac{v_0^2}{\sigma_0^2}=\lp 1-\delta_0\rp\frac{|\maw_{xx}^{\rm N}|}{|\maw_{zz}^{\rm N}|}-1+\frac{1}{3}\frac{\Lambda \mai_{xx}}{M\sigma_0^2}
\left[\frac{\mai_{zz}}{\mai_{xx}}\frac{|\maw_{xx}^{\rm N}|}{|\maw_{zz}^{\rm N}|}-1\right].
\end{equation}
Note that if $\Lambda$ cannot be neglected $v_0^2/\sigma_0^2$ is not only a function of the eccentricity for 
ellipsoids with the density give in (\ref{density1}), but depends also on the details of the matter density as the latter does not cancel. \\
\noindent Using almost the same set-up as above, we can use equation (\ref{curious}) to establish a relation between$v_0$, the mass $M$ and the geometry of the object which we choose below to be oblate. After straightforward
algebra we obtain
\begin{equation} \label{curious2}
\frac{v_0^2}{2\sigma_{0}^{2}}=\frac{3}{10}\frac{G_NM}{\sigma_{0}^{2}}\frac{\sqrt{1-e^{2}}}{a_{1}^{3}e^{2}}\left[\frac{3\arcsin e}{e}-\frac{3-e^{2}}{\sqrt{1-e^{2}}}\right]+(1-\delta_{0})\lp\frac{a_{1}}{a_{3}}\rp^{2}-1.
\end{equation}
Since this relation is derived from (\ref{curious}) which in turn is based on the assumption
of $\Lambda \neq 0$ it is only valid for non-zero cosmological constant albeit the latter does not enter
the expression. Note the enhancement factor $(a_{1}/a_{3})^2$. 
\begin{figure}
\begin{center}
\label{fomega}
\includegraphics[angle=270,width=10cm]{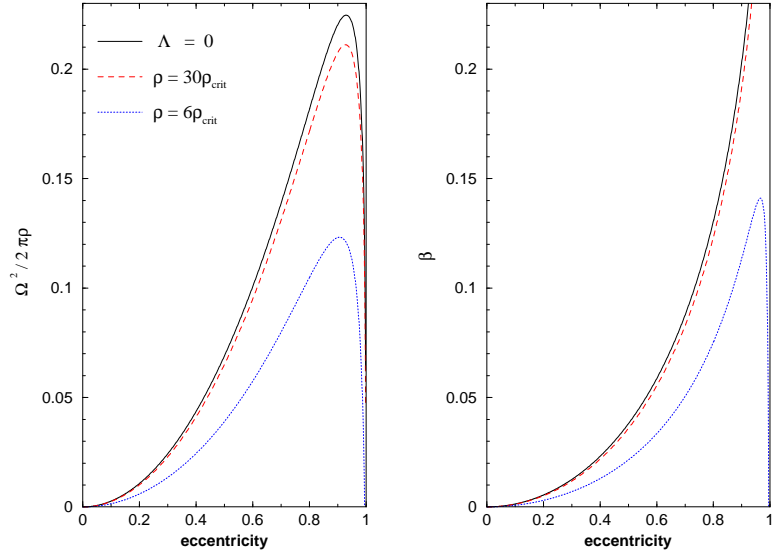}
\end{center}
\caption{\footnotesize{Effects of $\Lambda$ on the angular velocity 
$\Omega_{\rm rot}^{2}$ and the ratio $\beta$ for $\rho_{\rm vac}\approx 0.7\rho_{\rm crit}$ }}
\end{figure}
\subsection{Other effects of $\Lambda$ for non-rotating configurations}
We can now derive other relevant quantities from the scalar $\Lambda$-virial theorem applied 
to homogeneous ellipsoidal configurations. In this section we will not consider rotating configurations, 
but systems with kinetic energy coming from internal motions. We will focus again on ellipsoidal, oblate and prolate, geometries. 
As in the
preceding section the relevant quantity here is the function $\maa$  written for both configurations in equation (\ref{A}). 

\noindent \emph{Critical mass}: Consider the stability criteria for a homogeneous cloud with mass 
$M$ and internal mean velocity $\langle v \rangle$ in mechanical equilibrium with the background with pressure $P$ 
(see \cite{Padma2}). 
The system will collapse under it's own gravity ($P<0$) if its mass is greater than a critical mass $M_{c}$. 
With cosmological constant, this critical mass is increased with respect to its value $M_{c0}$ when $\Lambda=0$, which is expected, 
since now there is an \emph{external force} associated to $\Lambda$ that acts against Newtonian gravity 
and hence the collapse can be postponed. By using the scalar virial theorem (\ref{vt}) and setting $P=0$ 
as the criterion for the onset of instability, we can write for arbitrary geometry \\
\begin{equation}
\label{cmas}
M_{c}=M_{c0}\left[1-\maa \lp\frac{\rho_{\rm vac}}{\rho_{c0}}\rp\right]^{-1},\hspace{0.8cm}M_{c0}
=\frac{2\langle v^{2}\rangle}{|\tilde{\maw}^{\rm N}|},\hspace{0.8cm}\rho_{c0}=\frac{M_{c0}}{V}.
\end{equation} 
This expression is valid for any geometry. 
However, we pointed out already that spherical symmetry implies $\maa=2$ 
and the effect is suppressed. Some numerical values can be considered 
by writing $\langle v^{2}\rangle=3k_{B}T/m_{p}$ for a hydrogen cloud with $T\approx 500$ K and radius $R\approx 10$pc. One then has
\begin{equation}
\label{spt}
\maa\lp \frac{\rho_{\rm vac}}{\rho_{c0}}\rp \approx 10^{-8},
\end{equation} 
which represents a very tiny correction to the critical mass $M_{c0}$ for a spherically symmetric object. 
On the other hand, for ellipsoidal configurations with the same temperature, we have 
\begin{equation}
\label{ellt}
\delta M=\maa_{\rm obl} \lp\frac{\rho_{\rm vac}}{\rho_{c0}}\rp \approx 5\times 10^{-5} \left(\frac{a}{{\rm kpc}}\right)^{2},
\hspace{0.5cm}
\delta M=\maa_{\rm pro} \lp\frac{\rho_{\rm vac}}{\rho_{c0}}\rp \approx 10^{-1}\left(\frac{c}{{\rm kpc}}\right)^{2},
\end{equation} 
in the $e\to 1$ approximation for oblate and prolate configurations, respectively.
We have set $c=10a$ for the prolate case and $\Omega_{\rm vac}=0.7$. 
For an oblate ellipsoid with $a\approx 50$ kpc, one has $\delta M\approx 0.15$
while for the prolate ellipsoid with $c\approx 50$, $\delta M\approx 10^{2}$.

\noindent \emph{Mean velocity and Mass-Temperature relation}. 
By virtue of the scalar $\Lambda$-virial theorem, we can also write down the mass-temperature relation for an astrophysical structure.
Note first that  using $\mak=\frac{1}{2}M\langle v^{2}\rangle$ in equation (\ref{vt}) we have for $\Lambda=0$ the standard expression for
the mean velocity 
\begin{equation} \label{mean}
\langle v^2 \rangle_{\Lambda=0}=\frac{|{\cal W}^N|}{M}=\frac{\rho^2 G_N |\tilde{\maw}^N|}{M},
\end{equation}
where the second equality applies to the constant density case. Clearly, with $\Lambda=0$
the mean velocity cannot become zero. Let us
contrast it to the case with $\Lambda >0$. One obtains
\begin{equation} 
\label{mvel}
\langle v^{2} \rangle =\frac{\rho^{2}G_{\rm N}|\tilde{\maw}^{\rm N}|}{2M}\left[1-\maa \lp\frac{\rho_{\rm vac}}{\rho}\rp\right].
\end{equation}

As can been seen from (\ref{mvel}) the mean velocity in the ellipsoidal configurations is decreased because of the 
$\Lambda$-\emph{external force}. A drastic effect of the cosmological constant could be reached 
for the geometrical factor $\maa$ approaching the critical value $\maa_{\rm crit}=\rho/\rho_{\rm vac}$ 
which is possible for very flat objects. In the extreme the mean velocity can go to zero \footnote{In the case of angular velocity, i.e Eq (\ref{av}) this is somewhat different since $\Omega_{0}\to 0 $ as $e\to 1$ independently of the density.}
Together with inequality (\ref{ineq}), the result in Eq. (\ref{mvel}) tells us that the temperature , $T\propto \langle v^{2} \rangle$ of the objects is very small if the objects is above the limit to reach equilibrium (the square bracket in Eq. (\ref{mvel}) is then very small). This is a qualitative conclusion  based in the presence of non-zero cosmological constant. For ellipsoidal configurations, this defines also a maximum value for the eccentricity $e_{\rm max}$ given $\rho$ or vice-versa, i.e, a minimum value for the density $\rho_{\rm min}$ given $e$, through the relation $\rho_{\rm min}=\maa(e_{\rm max})\rho_{\rm vac}$. The bahavior of $e_{\rm max}$ is shown in Fig. 5 as a function of $\rho_{\rm vac}/\rho$. Galactic clusters with  $\rho_{\rm vac}/\rho\approx 0.1$ may have a vanishing mean velocity for $e_{\rm max}\to1$ in the oblate case and $e_{\rm max}\approx 0.92$ in the prolate case. If the density is smaller, a vanishing mean velocity can be reached for non so flat objects. Spherical configurations have $\langle v^{2} \rangle\to 0$ for $\rho\approx 2\rho_{\rm vac}$. \\
Since the mean velocity is proportional to the temperature, $\rho=M/V$ , Eq. (\ref{mvel}) represents also a mass-temperature relation. Hence the results for the mean velocity squared are also applicable to the temperature of the configuration.
For instance, we can write Eq. (\ref{mean}) for a cosmological structure, say a galactic cluster, 
by writing its density as resulting from a perturbation $\delta \rho$ from the background density of the universe 
$\rho_{\rm b}(t)$ as $\rho_{\rm gc}=\rho_{\rm b}(t)(1+\delta(t))$, where $\delta(t)=\delta\rho(t)/\rho_{\rm b}(t)$.  
Equation (\ref{mvel}) allows us to determine the temperature of the cluster at a given cosmic time $t$ as
\begin{eqnarray} 
\label{mvel2}
T&=&\frac{m_{p}}{10k_{B}}(1+\delta(t))a(t)^{-3}\Omega_{\rm b}H_{0}^{2}\mathcal{F}(e)\left[1-\maa(e)
\lp\frac{\Omega_{\rm vac}}{1-\Omega_{\rm vac}}\rp\lp1+\delta(t)\rp^{-1}\right], \\ \nonumber
\mathcal{F}(e)&=&\begin{cases}
a_{1}^{2}e^{-1}\sqrt{1-e^{2}}\ln\lp\frac{1+e}{1-e}\rp & \text{Prolate},\\
a_{1}^{2}e^{-1}\sqrt{1-e^{2}}\arcsin e & \text{Oblate},
\end{cases}
\end{eqnarray}
where $a(t)$ is the scale factor and $\mu$ is the mass of average components of the cluster. Equation (\ref{mvel2}) assumes a flat universe $\Omega_{\rm matter}+\Omega_{\rm vac}=1$. This result is a generalization of
a result derived in \cite{Wang} valid for spherical 
geometry. In 
that case one recovers the typical mass temperature relation $T\propto M^{2/3}$ mantaining $\rho$ constant. 
Although this has the same dependence as in equation (\ref{temperature3}), 
the meaning is different since (\ref{mvel2}) for $\maa=2$ computes the temperature of a certain galactic cluster at some redshift given its mass while equation (\ref{temperature3}) is associated to a reversible process where a configuration passes from $T=0$ to some final $T_{\star}$ through states of virial equilibrium mantaining a constant mass.

\begin{figure}[t]
\begin{center}
\includegraphics[angle=270,width=10cm]{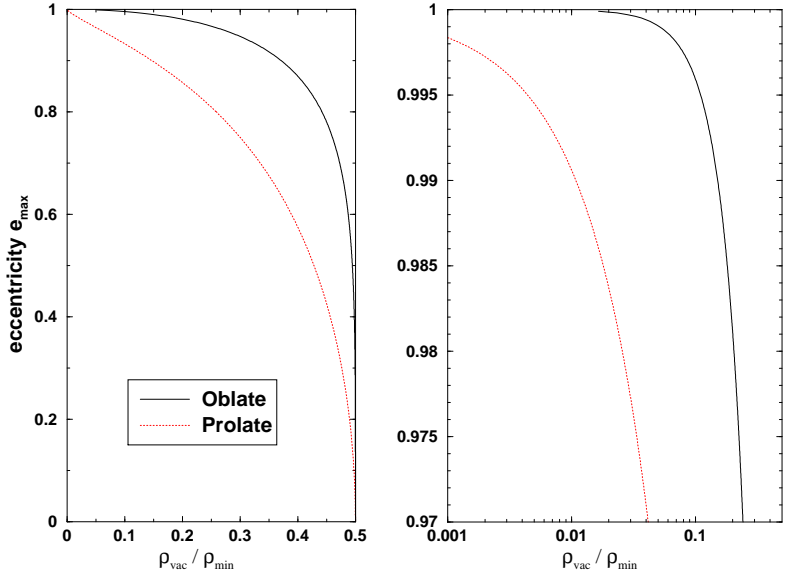}
\end{center}
\caption{\footnotesize{Behavior of $e_{\rm max}$ as a function of the ratio $\rho_{\rm vac}/\rho_{c}$ for oblate (solid line) and prolate (dashed line). In the left hand side we show the complete range, In the right hand side the same plot for $e\to 1$ and $\rho \ll \rho_{\rm vac}$}}
\end{figure}

\section{Small oscillations in the Newton-Hooke spacetime}
\begin{figure}
\begin{center}
\includegraphics[width=7cm,height=7cm]{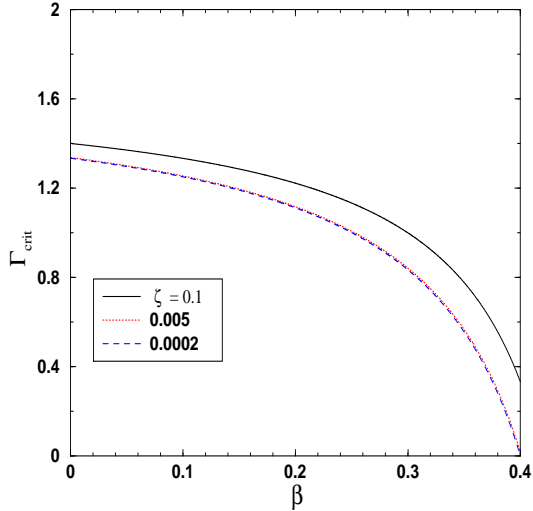}
\end{center}
\label{gamma}
\caption[]{\footnotesize{Critical adiabatic index as a function of 
$\beta$ for a homogeneous spherical configuration for different values of $\zeta=2\rho_{\rm vac}/\rho$}}
\end{figure}

The stability condition of Newtonian configurations against oscillations can be also derived from 
the second order virial equation by 
expanding the periodic Lagrangian perturbations $\xib(\vr,t)=\xib(\vr)e^{i\omega t}$ 
with oscillation frequency $\omega$. 
For simplicity, we will consider a rotational configuration without internal motions.
By assuming adiabatic perturbations, the variational form of the second order virial equation (\ref{virotro}) is written as (see \cite{Cha1,Friedman,Sha}) 
\begin{eqnarray}
\label{os21}
-\frac{1}{2}\lp \omega^{2}+\frac{16}{3}\pi \rho_{\rm vac}\rp \int_{V}\rho\lp r_{i}\xi_{k}+\xi_{i}r_{k}\rp \dtr&=& 
\int_{V}\rho\left[ \xi^{j}\paj \Phi^{\rm N}_{ik}+(1-\Gamma)\delta_{ik}\xi^{j}\paj \Phi^{\rm N}\right]\dtr \\ \nonumber
&+&\frac{1}{3}\delta_{ik}\int_{V}(5-3\Gamma)\rho\left[ (\om\cdot \xib)(\om\cdot \vr)
-\Omega^{2}(\vr \cdot \xib)\right]\dtr,
\end{eqnarray}
where $\Gamma=\lp\partial\ln P/\partial\ln \rho\rp_{s}$ is the adiabatic index governing the adiabatic perturbations. 
For polytropic e.o.s we have $\gamma=\Gamma$, of course.
The Lagrangian displacement is constrained with the boundary conditions $i)$, $\xib=0$ at $r=0$ and $ii)$ $\xib$ must be finite at the surface. 
The boundary conditions are required to solve the \emph{equation of motion} for $\xib$ 
resulting from perturbing Euler's equation (the Sturm-Liouville  eigenvalue equation).\\
Let us consider radial oscillations which are adequate for spherical systems and for small deviation from spherical symmetry. 
By assuming a general trial function for the Lagrangian displacement $\xib=f(r)\hat{\vr}$ 
satisfying the boundary conditions and taking the trace in Eq. (\ref{os21}), we can solve for frequency 
of the oscillation about equilibrium  
\begin{equation}
\label{os22}
\omega^{2}=\frac{8\pi \varrho}{A}\left[\Gamma-\frac{4}{3} -2\lp\Gamma-\frac{5}{3}\rp B 
-\frac{2}{3}A\lp\frac{\rho_{\rm vac}}{\varrho}\rp\right],
\end{equation}
where we have assumed  $\Gamma=$ constant throughout the configuration and we have defined in analogy to Eq. (\ref{ineq}) the quantities
\begin{equation}
\label{defi0}
|W|=\frac{1}{2}\varrho^{2}|\tilde{W}|,\hspace{0.5cm}I=\varrho \tilde{I},
\hspace{0.5cm}A\equiv \frac{16\pi}{3}\frac{\tilde{I}}{|\tilde{W}|}.
\end{equation}
In Eq. (\ref{defi0}) $\varrho$ is a parameter with units of density, and 
\begin{equation}
\label{defi1}
W\equiv -\int_{V}\rho(r)f(r)\frac{\dd \Phi^{\rm N}}{\dd r}\,\dtr,
\hspace{0.4cm}R\equiv \frac{1}{2}\Omega^{2}_{\rm rot}\int_{V}\rho(r) f(r)r^{-1}\left[
r^{2}-\delta_{ij}x_{i}x_{j}\right]\dtr,\hspace{0.4cm} I\equiv \int_{V}\rho(r) f(r)r\dtr,
\end{equation}
together with $B\equiv R/|W|$. The critical adiabatic index is written as
\begin{equation}
\label{ad1}
\Gamma_{\rm crit}=\frac{4}{3}\lp1-2B\rp^{-1}\left[1+\frac{1}{2}A\frac{\rho_{\rm vac}}{\varrho}-\frac{5}{2}B \right],
\end{equation}
such that for $\Gamma<\Gamma_{\rm crit}$ instability sets in and the system 
becomes unstable while perturbed. 
From Eq. (\ref{os22}), 
we see that through the inclusion of the cosmological constant 
we (the system) are forced to choose a bigger adiabatic index. 
In the simplest case, when $f(r)=r$, we have $W\to \maw$, $I\to \mai$, $B\to \beta$. 
Furthermore, for $\rho=$constant, we get $A\to \maa$ and $\varrho = \rho$. The stability condition then becomes 
\begin{equation}
\label{os24}
\Gamma>\frac{4}{3}\lp1-2\beta\rp^{-1}\left[1+\frac{1}{2}\maa\frac{\rho_{\rm vac}}{\rho}-\frac{5}{2}\beta \right].
\end{equation}
The spherical symmetry which we assumed compels us to write the ratio $\beta$ for low eccentricities 
or to parametrize it in terms of the total angular momentum of the configuration. 
In the first approximation, at low eccentricities the ratio $\beta$ given in (\ref{beta}) takes the form 
\begin{equation}
\label{betaap}
\beta\approx \frac{2}{3}\lp\frac{1}{5}-\frac{\rho_{\rm vac}}{\rho} \rp e^{2}+\mathcal{O}(e^{4}).
\end{equation}
This equation is useful if we want to calculate a small, but peculiar effect. Insisting that
$\Gamma$ is very close to $4/3$ we can convert the stability condition $\Gamma > \Gamma_{\rm crit}$
into a condition on eccentricity, namely 
\begin{equation}
\label{econ}
e>e_{\rm min}\equiv3.35 \lp1+5 \frac{\rho_{\rm vac}}{\rho}-3\varepsilon \rp^{1/2}\lp\varepsilon+\frac{4}{3}\frac{\rho_{\rm vac}}{\rho}\rp^{1/2},
\end{equation}
where $\varepsilon\equiv\frac{4}{3}-\Gamma$ measures a small departure from the critical value $4/3$. 
For the specific case $\varepsilon=0$, we conclude that the eccentricity must be such that $e_{\rm min}< e \ll1$ with 
\begin{equation}
\label{econ}
e_{\rm min}=3.8\lp\frac{\rho_{\rm vac}}{\rho}\rp^{1/2},
\end{equation}
which is clearly valid only for large densities, say for $\rho \ge  10^{3}\rho_{\rm vac  }$. 
This imposes a range for the eccentricity $e_{\rm crit}<e\ll 1$ for stability under radial perturbations.\\
For completely spherical configurations with constant density, we can write $\beta=\Omega_{\rm rot}^{2}/4\pi \rho$. 
Figure \ref{gamma} shows the behavior of the critical adiabatic index as 
a function of $\beta$ for different values of $\zeta=2\rho_{\rm vac}/\rho$. The largest deviation is for low densities as
expected. 
\\
For non rotating configurations, the stability criteria is simply given by
\begin{equation}
\label{stac2}
\Gamma_{\rm crit}=\frac{4}{3}\left[1+\frac{1}{2}A\frac{\rho_{\rm vac}}{\varrho}\right],\hspace{0.5cm}A=\frac{8\pi}{3}\varrho\frac{\int_{0}^{R} \rho(r)r^{2}f(r)\dd r}{\int_{0}^{R}\rho(r)r^{2}f(r)\frac{\dd \Phi^{\rm N}}{\dd r}\dd r}.
\end{equation}
If we work with a constant density profile, 
the choice of another trial function satisfying the boundary conditions mentioned before 
will not play a role in the above expressions since $A\to \maa=2$ for any $f(r)$.\\

\noindent Related results regarding the adiabatic index in cosmologies with non-zero $\Lambda$
have been obtained in the relativistic frame work in \cite{BH}.

\section{Conclusions}
In the present work we investigated in detail the astrophysical relevance of the 
cosmological constant for equilibrium configurations. Using the tensor and scalar virial equations and the
Lane-Emden equation we could show that many astrophysical facets get modified by $\Lambda$.
The second aspect concerns the fact that $\Lambda$ introduces new relevant scales (these scales
would be zero if $\Lambda=0$) like the maximal
virial volume defined by the maximal virial radius (\ref{rvir2}) and the maximal extension of bound orbits given
in Eq. (\ref{sca3}).  \\
It is often assumed that superclusters with densities $\sim \rho_{\rm crit}$ are not in equilibrium. With the inequality (\ref{ineq}) we have a precise tool to quantify this statement. Indeed, the pancake structure of the superclusters lead us to the conclusionthat they are even far away from the equilibrium state due to the factor $\maa$ which grows with the object's flatness. On the other hand, relatively low density objects can still reach equilibrium (even if $\maa=2$ for the spherical case, the density of the object has to be only twice as large as $\rho_{\rm vac}$, i.e, $1.4\rho_{\rm crit}$). There is nothing which could, in pronciple, prevent a relatively low density object to be in equilibrium This is not only a result obtained from the virial equations with $\Lambda$, but follows also from hydrostatic equilibrium with the inclusion of $\Lambda$ \cite{Bala1}. It is then natural to put forward the question how the shape of such objects affect their properties as compared to the case without $\Lambda$. For instance, if $\rho/\rho_{\rm vac}=10$ and $e\sim 0.9$, the angular velocity is reduced approximately $30\%$ with respect to its value with $\Lambda=0$ ($\Omega_{\rm rot}=0.7\Omega_{0}$). Other effects discussed in the text are the mean velocity (see Fig. 5) and the adiabatic index (see Fig. 6). To be specific let us takje an example of a very low density object, $\rho=2.566 \rho_{\rm crit}$. This object is still virialized as long as its eccentricity is $e<0.5$. Approaching $e\to 0.5$ the mean velocity (temperature) goes to zero until finally beyond $0.5$ the objects ceases to be in equilibrium.\\

To highlight  some quantities which get affected let us mention the polytropic index $n$ and the angular velocity.
Low-density objects with $\rho \sim  20 \rho_{\rm crit}$ an index $n > 1.5$ 
becomes unphysical as the the conglomeration of matter does not have a definite finite radius. Taken together
with the results of section five on stability against small oscillations this implies that
the adiabatic index $\gamma$ from the equation of state $P \propto \rho^{\gamma}$ gets 
restricted for these densities in a rather strong way, namely by $1.33 < \gamma < 1.66$.
For higher densities the effect grows with $n$. For instance, for $\rho \sim 10^4 \rho_{\rm crit}$, 
$n=5$ is an unacceptable solution due to the lack of a finite extension. 
For non-spherical configurations we could show an effect of $\Lambda$ on angular velocities, 
the mean velocity of the component of the astrophysical object and on the critical mass, again
for densities one and two orders of magnitude above the critical one. We think that
the work can be generalized in various ways. For instance, to generalize
the virial equations to be able to differentiate between different models 
of dark energy mentioned at the beginning. Another aspect is to look into 
the Lane-Emden equation for non-spherical geometries \cite{Bala1,Lombardi} 
. Here we concentrated on the equilibrium condition of already virialized matter. The collapse
of matter in an accelerated universe with dark energy has been considered in \cite{Mota}.

\section*{Appendix}
Here we briefly show the solution for the cubic equation (\ref{cubic}). Let us write that expression as
\begin{equation}
\label{co4}
y^{3}+py+q=0,\hspace{0.5cm}p=10\eta\rl^{2}>0,\hspace{0.5cm}q=-3\rs \rl^{2}<0,
\end{equation}
corresponding to a positive cosmological constant. Associated with this expression one defines a  discriminant and a quantity $R$ defined as
\begin{equation}
\label{co5}
D=\frac{1}{27}p^{3}+\frac{1}{4}q^{2}>0,\hspace{0.8cm}R=\text{sgn}(q)\sqrt{\frac{1}{3}|p|}<0.
\end{equation}
Since $p>0$, the roots of (\ref{co4}) are given in term of the auxiliary angle $\phi$ defined as 
\begin{equation}
\label{co6}
\sinh \phi =\frac{q}{2R^{3}},
\end{equation}
so that the only real solution for (\ref{co4}) is written as
\begin{equation}
\label{co7}
y=-2R\sinh 
\lp\frac{\phi}{3}\rp=\sqrt{\frac{4}{3}|p|}\sinh\left[\frac{1}{3}{\rm arcsinh}\lp\frac{q}{2R^{3}}\rp\right].
\end{equation}
The  other two roots are complex numbers whose real part are negative. If $p<0$ and $q>0$ the roots depend whether $D$ is bigger or smaller than zero. If $D\leq0$, the three roots are real and negative, while $D>0$ yields one real negative root and two complex roots which real parts are given as
\begin{equation}
\label{co8}
y=\sqrt{\frac{1}{3}|p|}\cosh\left[\frac{1}{3}{\rm 
arccosh}\lp\frac{q}{2R^{3}}\rp\right].
\end{equation}

\end{document}